\documentclass[12pt]{article}

\usepackage{amsmath,amsfonts,amssymb,amsthm}
\usepackage{color, graphicx}
\usepackage{epsfig,fullpage}
\usepackage{natbib}
\usepackage{url}
\usepackage{lineno}
\usepackage{multirow}
\usepackage{subfig}
\usepackage{rotating,adjustbox}
\usepackage{subfig}
\usepackage{hyperref}

\newcommand{\BEAS}{\begin{eqnarray*}}
\newcommand{\EEAS}{\end{eqnarray*}}
\newcommand{\BEA}{\begin{eqnarray}}
\newcommand{\EEA}{\end{eqnarray}}
\newcommand{\BIT}{\begin{itemize}}
\newcommand{\EIT}{\end{itemize}}
\newcommand{\BNUM}{\begin{enumerate}}
\newcommand{\ENUM}{\end{enumerate}}

\newcommand{\Expect}{\mathop{\bf E{}}}

\begin{document}

\title{Covariate-dependent control limits for the detection of abnormal price changes in scanner data}

\author{Youngrae Kim, Sangkyun Kim, Johan Lim, Sungim Lee, \\ Won Son, and Heejin Hwang\thanks{Youngrae Kim and Johan Lim are at the Department of Statistics, Seoul National University, Seoul, Korea. SangKyun Kim and Sungim Lee are at the Department of Applied Statistics, Dankook University, Yongin, Korea. Won Son and Heejin Hwang are at the Bank of Korea, Seoul, Korea. All correspondence should be directed to Sungim Lee, e-mail:\texttt{silee@dankook.ac.kr}.} }

\date{}
\maketitle
\begin{abstract}
\noindent
Currently, large-scale sales data for consumer goods, called scanner data, are obtained by scanning the bar codes of individual products at the points of sale of retail outlets. Many national statistical offices use scanner data to build consumer price statistics. In this process, as in other statistical procedures, the detection of abnormal transactions in sales prices is an important step in the analysis. Popular methods for conducting such outlier detection are the quartile method, the Hidiroglou-Berthelot method, the resistant fences method, and the Tukey algorithm. These methods are based solely on information about price changes and not on any of the other covariates (e.g., sales volume or types of retail shops) that are also available from scanner data. In this paper, we propose a new method to detect abnormal price changes that takes into account an additional covariate, namely, sales volume. We assume that the variance of the log of the price change is a smooth function of the sales volume and estimate the function from previously observed data. We numerically show the advantages of the new method over existing methods. We also apply the methods to real scanner data collected at weekly intervals by the Korean Chamber of Commerce and Industry between 2013 and 2014 and compare their performance. 

\noindent

\end{abstract}

\baselineskip 18pt

\section{Introduction}

In the analysis of large-scale data, preprocessing steps, which include outlier detection and normalization, are
required. These steps are especially important when the data are used in real-world applications such as the creation of consumer price statistics. In particular, outliers or anomalies introduce bias in the analysis and should therefore be removed. In this paper, we discuss the procedure to detect outliers in scanner data.

Scanner data are detailed data on product sales obtained by scanning the bar codes of individual products at the points of sale of retail outlets. The data contain information about which items are sold at which stores, along with the volumes and prices of the sold items. Both researchers and practitioners are interested in the use of these data and make efforts to analyze them. In particular, many national statistical offices (NSOs), including those of the UK, Switzerland, and the Netherlands, use scanner data when calculating various price indices \citep{Bird:2014,deHaan:2011,deHaan:2014}. The goal of our research is to detect abnormal transactions in prices from scanner data, as the existence of anomalies in the data results in biased conclusions about price trends and variations.

Outlier detection in price changes is often linked to the traditional consumer price index (CPI) survey \citep{Saidi:2005,Rais:2008}. There are four representative methods for outlier detection: the quartile method, the Hidiroglou-Berthelot method, the resistant fences method, and the Tukey algorithm. \citet{Saidi:2005} at Statistics Canada and \citet{Rais:2008} show that the quartile method is preferred over other methods, while the UK NSO uses the Tukey algorithm to detect outliers in its CPI. The \citet{ILO:2004} mentions the quartile method, the Hidiroglou-Berthelot method and the Tukey algorithm but does not favor one method over another as a tool for outlier detection. In addition, \citet{Thompson:1999} from the United States Bureau of the Census (USBC) recommend the resistant fences method for use on price index data. These four methods construct intervals for price ratios and mark price changes that do not fall within the intervals as possible errors or outliers. Unlike traditional CPI survey data, scanner data have information not only on prices but also on other sales characteristics, such as the sales time and volume of each item. For this reason, the traditional methods used to process CPI survey data, which are only based on price information, are not well suited for use with scanner data.

\begin{figure}[htb!]
\centering
\includegraphics[width=10cm, height=8cm]{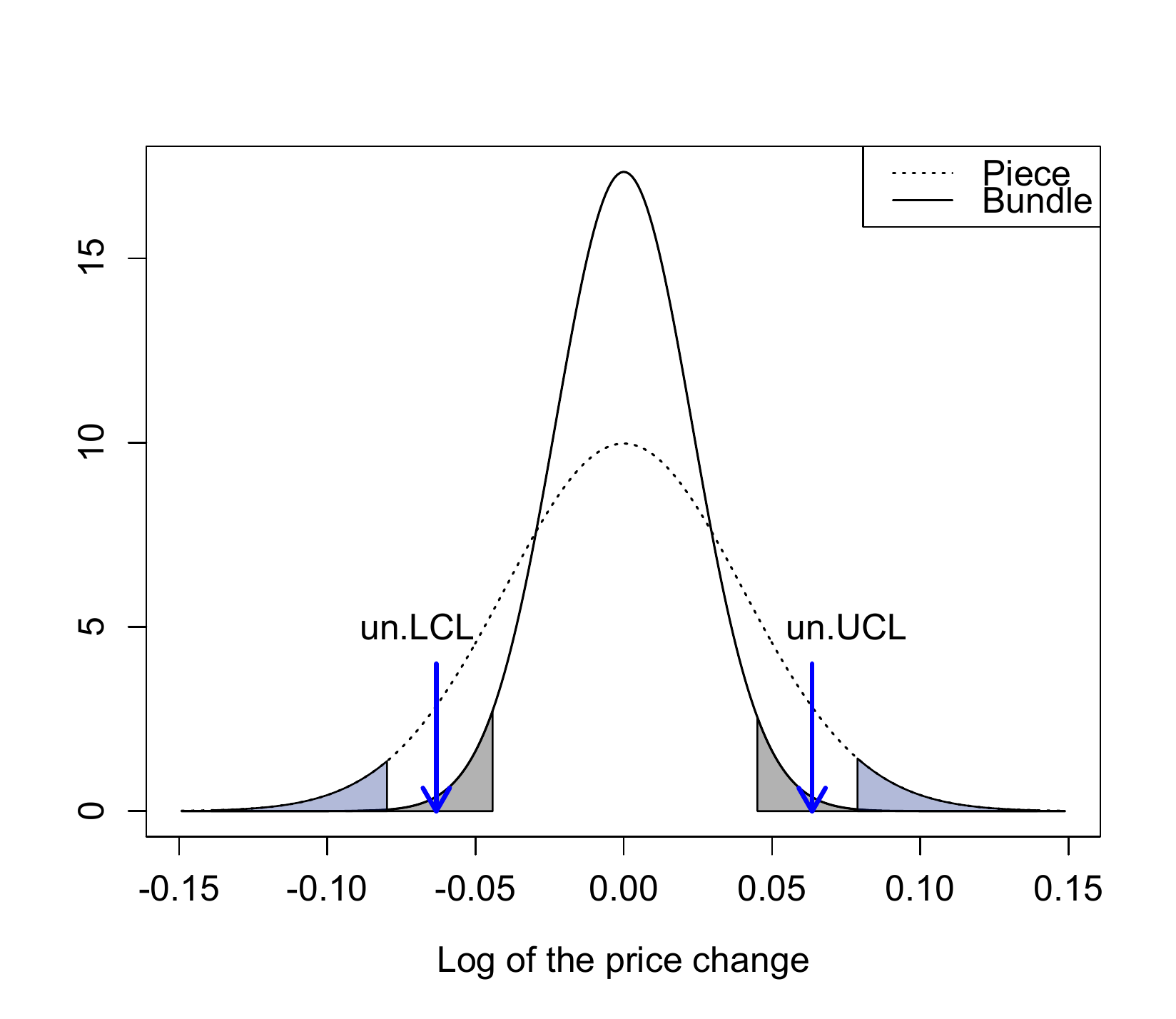}
\caption{Distribution of log of the price change: the dotted line indicates cases in which an item is sold by the piece, $N(0, \sigma_1^2)$, and the solid line indicates cases in which an item is sold in a bundle $N(0, \sigma_2^2)$. In each histogram, the 5\% two-sided tail area is shaded. The arrows are the unconditional lower (un. LCL) and upper (un. UCL) 95\% control limits of the log of the price change evaluated without using the volume information.}
\label{fig1:toy}
\end{figure}

To understand the difficulties of using traditional methods to assess scanner data, let us consider an example where the variation of the log of the price change (see the beginning of Section 2) depends on sales volume.
Suppose the price ratio of an item follows the lognormal distribution $\log N (0, \sigma_1^2)$ if it is sold by the piece, whereas
it follows the lognormal distribution $\log N (0, \sigma_2^2)$ ($ \sigma_2^2 \le \sigma_1^2$) if it is sold in a bundle.
For this item, 20\% of the sales volume is sold by the piece, whereas 80\% is sold in a bundle. However, we only
have information on the price, and no information on the sales volume is available. Then, the observed log of
the price ratio is a mixture of two normal distributions, $N (0, \sigma_1^2)$ and $N (0, \sigma_2^2)$.
The intervals constructed by the traditional methods lie between the intervals for the piece sales and the bundle
sales, as shown in Figure \ref{fig1:toy}. This shows that traditional methods that do not use volume information falsely identify many normal piece sales as abnormal and incorrectly judge many abnormal bundle sales as normal.

To resolve this problem, we propose a new method for detecting outliers in scanner data. As in the traditional methods, we set the confidence limits for the log of the price change and define price changes that exceed these limits as outliers. However, unlike the traditional methods, where the values of the limits are constant on sales volume, we allow the upper and lower limits of the intervals to depend on sales volume, our additional covariate. To do so, we assume that the variance of the log of the price change is a smooth function of sales volumes and estimate the function from previously observed data.

The rest of this paper is organized as follows. In Section 2, we review the existing outlier detection methods. In Section 3, we introduce a procedure to estimate the variance function of the log of the price change and propose a new outlier detection method for use on scanner data. In Section 4, we numerically show the advantages of the new method over existing methods. In Section 5, we apply the methods to the scanner data collected in weekly intervals by the Korean Chamber of Commerce and Industry (KCCI) between 2013 and 2014. We conclude the paper with a brief summary and discussion in Section 6.

\section{Outlier detection methods for scanner data}

When the unit price at time $t$ for an item is set as $P_t$, the price change between time $t$ and $t-1$ is defined as the ratio $R_t = P_t/P_{t-1}$. The four existing methods---the quartile, Hidiroglou-Berthelot, resistant fences, and Tukey algorithm methods---construct an interval for $R_t$ (or $\log R_t$) and define $R_t$ as an outlier if it falls outside the interval. This corresponds to the concept of statistical quality control, which motivates our interest in finding the proper interval for monitoring price change $R_t$ (or $\log R_t$). In the rest of this paper, we refer to this interval as being defined by control limits. Hence, control charts visualize the control limits of normal price change variation.

\subsection{Quartile method (Quartile)}

The quartile method is regarded as the primary outlier detection method in a conventional sense. It builds the control limits using quartiles for $R_t$ as in a standard boxplot \citep{Tukey:1977}. Let $Q_1, Q_2,$ and $Q_3$ be the first, second, and third quartiles, respectively. Then, the upper control limit (UCL) and the lower control limit (LCL) for the price change are defined as follows:
\begin{equation}
\begin{aligned}
 {\rm UCL}_{QM} &= Q_2+c_u  (Q_3-Q_2)\\ 
 {\rm LCL}_{QM}  &= Q_2- c_l  (Q_2-Q_1) , 
\end{aligned} \label{qm}
\end{equation}
where $c_u$ and $c_l$ are typically set as the same and predetermined to take into account the distributional characteristics
of $R_t$. If $R_t$ follows the normal distribution, $Q_2$ is equal to $\mu$ and $\Expect[IQR]=2 \Phi^{-1}(0.75) \sigma  \approx 1.349 \sigma$. Hence, $c_u=c_l=4.5$ approximately leads to the interval
$( \Expect[{\rm LCL}_{QM}],\Expect[{\rm UCL}_{QM}])=(\mu - 3 \sigma, \mu + 3 \sigma)$, whose probability of type I error under normality is 0.27\%.


One difficulty of the quartile method comes from the local constancy of
$R_t$ over $t$. This gives many observations of $R_t$ a value of 1, bringing the quartiles
closer to each other ($Q_1\approx Q_2\approx Q_3$), and further
resulting in many false outlier detections. (This problem is addressed in the Tukey algorithm through the use of Tukey samples, which we introduce in Section 2.4.) 
The second difficulty of the quartile method arises from the skewed data.
Supposing that the data are right skewed and $c_u=c_l$, the upper control limit is
more variable and sensitive to changes in the training data than the lower control
limit. We may adjust $c_u$ and $c_l$ separately to resolve this difficulty. However,
in practice, it is not an easy task to adjust them together
to achieve the given level of confidence while keeping the two limits equally sensitive. For this reason,
\citet{Saidi:2005} and \citet{Thompson:1999} suggest using the log transformation of $R_t$ for the analysis.

\color{black}

\subsection{Hidiroglou-Berthelot (HB) method}

\citet{Hidi:1986} introduce a new transformation of the price ratios, noting that price changes are unequal in representing price decreases and increases. For example, if an item is offered at half price, the price is reduced by 50 percent. Returning to the original price implies a 100 percent price increase. To compensate for this asymmetry in price changes, the authors propose the following transformation, called the HB transformation:
\begin{equation}
S_t = \left\{ \begin{array}{l@{\quad:\quad}l}
         1 -\frac{Q_2}{R_t} & 0<R_t <Q_2\\  
         \frac{Q_2}{R_t} -1 & R_t \ge Q_2 \, , \label{HB} 
         \end{array} \right.
\end{equation}
where $Q_2$ stands for the median of $R_t$. This transforms price ratios less than $Q_2$ into negative values and price ratios greater than $Q_2$ into positive values. The authors suggest applying the quartile method to the HB-transformed data, $S_t$. Suppose $R_t$ is equal to one for almost $t$, i.e., if the price changes are minor, the HB transformation in (\ref{HB}) becomes the first-order Taylor expansion of log $R_t$, that is, $S_t \approx \log R_t$. The quartile method with log-transformed data and the HB method are similar to each other.

\subsection{Resistant fences (RF) method}

The RF method defines the control limits using interquartile range \citep{Thompson:1999}. The upper control limit (UCL) and the lower control limit (LCL) for the price change ($R_t$) are given as follows:
\begin{equation}
\begin{aligned}
 {\rm UCL}_{RF} & = Q_3 +c_u  (Q_3-Q_1)\\
 {\rm LCL}_{RF}  & = Q_1 - c_l  (Q_3-Q_1)\,,
\end{aligned} \label{RF}
\end{equation}
where $c_u$ and $c_l$ are commonly set as the same predetermined constant $c$. The constant $c$ is usually set to 1.5 (inner fence) and 3 (outer fence). The difference between the two values in (\ref{RF}) becomes $(2c+1) (Q_3-Q_1)$. Hence, $c=1.75$ (\ref{RF}) results in an interval length of approximately 6$\sigma$ as in the quartile method setting $c=4.5$, where $R_t$ follows the normal distribution. Note that these two intervals approximately coincide under normality:
\begin{equation}
\begin{aligned}
 {\rm LCL}_{QM} & = Q_2 -4.5 (Q_2 - Q_1)\\
                            & = Q_1 + (Q_3-Q_1)/2 -4.5 (Q_3 - Q_1)/2\\
                            &=  Q_1 - 1.75(Q_3-Q_1) = {\rm LCL}_{RF} .\\
\end{aligned} \label{QR}
\end{equation}
Similarly, the UCLs of the quartile and RF methods are also the same.

\subsection{Tukey algorithm (Tukey)}

As mentioned earlier, the Tukey algorithm is popularly used by the UK NSO to detect outliers in CPI data. Unlike the quartile method, it works well even if the variability of the price is small. The Tukey algorithm removes data with no change in the price ($R_t=1$) and uses only the observations of $R_t$ not equal to $1$, which are called Tukey samples. Suppose $\{R_1^s, R_2^s, \cdots , R_n^s\}$ are Tukey samples. Then, the control limits are defined as
\begin{equation}
\begin{aligned}
{\rm UCL}_{TA} & = \bar{R}^s + 2.5(\bar{R}_U^s - \bar{R}^s)\\
{\rm LCL}_{TA} & = \bar{R}^s - 2.5(\bar{R}^s - \bar{R}_L^s),
\end{aligned} \label{ta}
\end{equation}
where $\bar{R}^s$ is the sample mean of the Tukey samples, $\bar{R}_U^s$ is the average of the Tukey samples larger than $\bar{R}^s$, and $\bar{R}_L^s$ is the average of the Tukey samples that are smaller than $\bar{R}^s$. The constant
$2.5$ in (\ref{ta}) is the value to satisfy $\Expect[ {\rm UCL}_{TA}]  =  \mu + 2.5 \sqrt{\frac{2}{\pi}} \sigma \approx \mu+2\sigma$ and $\Expect[ {\rm LCL}_{TA}] = \mu - 2.5 \sqrt{\frac{2}{\pi}} \sigma \approx \mu-2\sigma$ under normality on $R^s_t$.
Here, we remark that the parameters $(\mu, \sigma^2)$ are for the Tukey samples and not for the original price ratios. A small number of Tukey samples may not provide accurate outlier detection \citep{Rais:2008} because only a fraction of the data is used.

\section{Covariate-dependent control chart}

As mentioned in the Introduction, there are some difficulties in applying existing methods such as the quartile, HB, and RF methods and the Tukey algorithm to scanner data. These methods all consider the price only when determining control limits for price changes. We expect that the existing methods will yield higher type I errors or more false positives if the variance of price changes depends on the quantity sold. Therefore, we propose a new method to improve on the control limits in previous methods by incorporating additional information, that is, sales volume data. Here, we assume that the log of the price change $Y_t=\log(R_t)$, for $t=2,3,\ldots,T$, follows the model
\begin{equation}\label{eqn:model}
Y_t =\mu(V_{t-1}, V_t) +\sigma(V_{t-1}, V_t) \epsilon_t,
\end{equation}
where $\mu(x_1, x_2)$ and $\sigma(x_1,x_2)$ are continuous functions in $x_1$ and $x_2$, and $\epsilon_t$ is IID from a distribution with the mean of $0$ and a variance of $1$. Additionally, $V_{t}$ and $V_{t-1}$ are the sales volumes at times $t$ and $t-1$, respectively. If the log of the price change is under the in-control status, there is no abrupt change in the price at time $t$, implying $\mu(V_{t-1}, V_t)=0$. The control limits at time $s$ become simply
\begin{equation} 
 \pm ~3  \cdot \sigma(V_{s-1}, V_{s}). \label{eqn:oracle-limit}
\end{equation}
Here, the control limits in (\ref{eqn:oracle-limit}) yield a type I error of 0.27\% when $Y_t$ is normally distributed. Moreover, if $\sigma(V_{s-1}, V_{s})$ is a constant function, it corresponds to the limits of the quartile for $Y_t$ with $c=4.5$.

The variance function $\sigma(x_1,x_2)$ is unknown in practice and needs to be estimated from the data. Many
methods are proposed to estimate the variance functions in the heteroscedastic regression model \eqref{eqn:model}.
Two of the major approaches are the residual-based and difference-based methods. The former estimates the variance function by estimating the mean of the squared residual
$r(x_1,  x_2) = (y - \mu(x_1,x_2) )^2$ and using $\Expect \left(r(x_1,x_2)\right)= \sigma^2(x_1,x_2)$. Here, under the in-control status, $\mu(x_1,x_2)=0$. In practice, the squared residuals are evaluated at the ``data points'' $(x_{i1},x_{i2},\widehat{r}_i)$, where $\widehat{r}_i = ( y_i - {\widehat \mu} (x_{i1},x_{i2}) )^2$ for $i=1,\dotsc,n$, and the mean function $\mu(x_{i1},x_{i2})$ is plugged in with its estimate $\widehat{\mu}(x_{i1},x_{i2})$. The local polynomial (e.g., linear, quadratic) regression estimator is widely used to estimate the mean of the squared residuals \citep{HC:89,FY:98}. For given $(x_1, x_2)$, it solves
\begin{equation} \label{eqn:local-poly}
 \big(\widehat{\alpha}, \widehat{\boldsymbol \beta} \big)  = \arg\min_{\alpha, {\boldsymbol \beta}} 
\sum_{i=1}^n \Big\{ \widehat{r}_i - \alpha - \sum_{j=1}^k \sum_{\ell_1,\ell_2}\beta_{j} (\ell_1,\ell_2) \big( x_{i1}-x_1 \big)^{\ell_1} \big( x_{i2}-x_2 \big)^{\ell_2} \Big\}^2  K_h \left( (x_1, x_2) (x_{i1}, x_{i2}) \right), 
\end{equation}
where ${\boldsymbol \beta}$ is the vector of $\big\{ \beta_j(\ell_1,\ell_2) |  \ell_1+\ell_2=j, \ell_1, \ell_2 \ge 0 , j=1,2,\ldots,k \big\}$,   $K_h \left( (x_1, x_2) (x_{i1}, x_{i2}) \right)=K \left( {(x_{i1}-x_1})/{h_1},  {(x_{i2}-x_2)}/{h_2}\right)$, and $ h=(h_1, h_2)$ with $h_1,h_2>0$ as the bandwidth. The residual-based variance function estimator is defined as $\widehat{\sigma}^2(x_{1},x_{2})=\widehat{\alpha}$.

The second popular procedure for the variance function is the difference-based method,
which actually does not require the estimation of the mean function. The method utilizes the fact that
when the data points are sorted such that
$x_1 \le x_2 \le \cdots \le x_n$, the pseudoresidual
\begin{equation}\nonumber 
\widehat{\sigma}^2 (x_i)= \left(\textstyle\sum_{j=-r}^r w_j y_{i+j} \right)^2,
\end{equation}
where $r>0$ is a fixed constant and the coefficients $\{w_j\}$ satisfy $\sum_{j=-r}^r w_j=0$ and $\sum_{j=-r}^r w_j^2=1$,
yielding an unbiased estimator of the variance $\sigma^2(x)\equiv\sigma^2$.
For example, if $r=1$, $w_1=1/\sqrt{2}$, $w_0=-w_1$, and $w_{-1}=0$, the estimator becomes $\widehat{\sigma}^2(x_i)= (y_{i+1} - y_{i})^2 /2$. In the heteroscedastic model (\ref{eqn:model}) above,
\citet{brown2007variance} consider applying the local linear regression to the pseudoresiduals to estimate the variance function $\sigma^2 (x)$. However, in our case, the variance function is bivariate, and the difference-based methods are not directly applicable.

\section{Simulation study}

In this section, we numerically investigate the performance of our proposal (Var) in Section 3 to detect abnormal changes in the price. We consider six methods for comparison: (i) the method with constant variance independent of sales volume (Const), (ii) the quartile method, (iii) the HB method (iv) the RF method, (v) the Tukey algorithm, and (vi) the method with a known true variance function (Oracle). The {\it Oracle} method is the gold standard, and existing methods such as quartile, HB, RF, and Tukey do not use sales volume information.

The data for the simulation study are generated as follows. In each dataset, we generate $600$ data points, where the first $300$ are under the control status (training period) and the next $300$ data points are to be tested (the period possibly having abnormal changes).
The data are generated from the following model for $t=1,2,\ldots,600$:
\begin{equation}\label{eqn:simu-model} 
P_t = P_{t-1}\exp \big(\sigma(V_{t-1}, V_t)\epsilon_t+\delta_t\big),
\end{equation}
where $P_0=1$, $V_t \overset{i.i.d.}{\sim} 1+\chi^2(5)$, $\epsilon_t \overset{i.i.d}{\sim} N(0,1)$, and $\delta_t= 2$ for $t\in J \subset \big\{301, 302, \ldots, 600 \big\}$, and otherwise, $\delta_t=0$.

We set the proportion of abnormal changes to comprise $5\%$ (15 data points) and $10\%$ (30 data points) of the testing data. Finally, we consider three cases regarding the variance function $\sigma^2(v_1, v_2\big)$: (a)  $\sigma^2(v_1, v_2) = 1$, (b)
$\sigma^2(v_1, v_2) = (1/46) v_1^2$, and (c) $\sigma^2(v_1,v_2) = ({1}/{92}) (v_1+v_2)^2$. Case (a) implies that the variance function does not depend on the sales volume. In case (b), the variance is affected only by the sales volume at time $t$. Finally, case (c) implies that the variance of day $t$ is complexly related to the sales volume at both time $t$ and $t-1$. Note that in cases (b) and (c), we adjust the expected value of the variance function $\sigma^2 (v_1, v_2)$ to be equal to one as in case (a). The constant $c$ for the quartile and HB methods is set to 4.5 and is set to 1.75 for the RF method to have the same false alarm rate of 0.27\%.

We compare the abovementioned seven methods, which include our approach, in terms of their sensitivity (SEN), specificity (SPE) and accuracy (ACC). The three measures are formally defined as
\begin{equation}\label{eqn:measures}
\begin{aligned}
&{\rm SEN} = \frac{\rm TP}{\rm P} = \frac{\rm TP}{{\rm TP}+{\rm FN}} \\
&{\rm SPE} = \frac{\rm TN}{\rm N} = \frac{\rm TN}{{\rm TN}+{\rm FP}} \\
&{\rm ACC} = \frac{{\rm TP}+{\rm TN}}{{\rm P}+{\rm N}} = \frac{{\rm TP}+{\rm TN}}{{\rm TP}+{\rm TN}+{\rm FP}+{\rm FN}}
\end{aligned},
\end{equation}
where true positive (TP) refers to the number of data points that are determined to be positive among observations of abrupt changes (P); true negative (TN) is the number of data points determined to be negative among the normal observations (N); false negative (FN) is the number of data points that are determined to be normal points among observations of abrupt changes; and finally, false positive (FP) is the number of data points determined to be positive among the normal observations.

We simulate $50$ datasets for the three types of variance functions and two choices of the proportion of abnormal changes. In each dataset, the variance function is estimated using the local constant regression, which is the solution to \eqref{eqn:local-poly}
with $\beta_j (\ell_1,\ell_2)=0$ for all $j$ and $(\ell_1, \ell_2)$ as
\begin{equation} \label{eqn:var-ftn-est}
\hat{\sigma}^2(v_1,v_2) = \hat{f}(v_1,v_2) = \frac{\sum_{i\in [T]} K_h \big((v_1,v_2), (V_{i-1}, V_i) \big) Y_i^2}{\sum_{i\in [T]}K_h \big((v_1,v_2), (V_{i-1}, V_i) \big)}
\end{equation}
where $[T]$ is a set of training data used for estimation and $K_h$ is the second-order Gaussian kernel function obtained by differentiating the Gaussian kernel twice. The kernel estimator (\ref{eqn:var-ftn-est}) is supported by the `\emph{npreg}' function of the `\emph{np}' package in the R software. The bandwidth $h$ is selected by a cross-validation procedure minimizing the cross-validated error ${\rm CVerr}$,
\begin{equation} \nonumber
{\rm CVerr} (h) = \sum_{i\in [T]}\big(Y_i^2-\widehat{f}_{[-i]}(V_{i-1}, V_i) \big)^2,
\end{equation}
where $\widehat{f}_{[-i]}(V_{i-1},V_i)$ is the kernel estimator calculated without the $i$-th observation. 

\begin{table} [htb!]
\begin{footnotesize} 
\begin{center} 
\begin{tabular}{|c|c|c|c|c|c|c|c|c|}
\hline
Variance & Method & TN & FN & FP & TP & SEN & SPE & ACC\\
\hline
\multirow{14}{*}{Case (a)} & \multirow{2}{*}{Var} & 269.76 & 24.10 & 0.88 & 4.26 & 0.15 & 1.00 & 0.92\\
 &  & (1.70) & (2.72) & (1.06) & (1.93) & (0.07) & (0.00) & (0.01)\\
\cline{2-9}
& \multirow{2}{*}{Const} & 270.00 & 24.16 & 0.64 & 4.20 & 0.15 & 1.00 & 0.92\\

 &  & (1.87) & (2.54) & (0.88) & (1.78) & (0.06) & (0.00) & (0.01)\\
\cline{2-9}
 & \multirow{2}{*}{Quartile} & 269.76 & 24.02 & 0.88 & 4.34 & 0.15 & 1.00 & 0.92\\

 &  & (2.08) & (2.77) & (1.19) & (1.97) & (0.07) & (0.00) & (0.01)\\
\cline{2-9}
 & \multirow{2}{*}{HB} & 244.82 & 9.98 & 25.82 & 18.38 & 0.65 & 0.90 & 0.88\\
 &  & (6.84) & (2.35) & (6.99) & (2.53) & (0.08) & (0.03) & (0.02)\\
\cline{2-9}
 & \multirow{2}{*}{RF} & 269.86 & 24.10 & 0.78 & 4.26 & 0.15 & 1.00 & 0.92\\
 &  & (2.04) & (2.93) & (1.02) & (2.12) & (0.08) & (0.00) & (0.01)\\
\cline{2-9}
 & \multirow{2}{*}{Tukey} & 257.80 & 13.64 & 12.84 & 14.72 & 0.52 & 0.95 & 0.91\\
 &  & (3.91) & (2.64) & (4.15) & (2.62) & (0.09) & (0.02) & (0.01)\\
\cline{2-9}
 & \multirow{2}{*}{Oracle} & 270.04 & 24.16 & 0.60 & 4.20 & 0.15 & 1.00 & 0.92\\
 &  & (1.84) & (2.47) & (0.83) & (1.65) & (0.06) & (0.00) & (0.01)\\
\hline
\multirow{14}{*}{Case (b)} & \multirow{2}{*}{Var} & 268.58 & 17.52 & 1.76 & 11.14 & 0.39 & 0.99 & 0.94\\

 &  & (2.72) & (2.86) & (1.76) & (3.11) & (0.10) & (0.01) & (0.01)\\
\cline{2-9}
 & \multirow{2}{*}{Const} & 266.00 & 25.38 & 4.34 & 3.28 & 0.11 & 0.98 & 0.90\\

 &  & (3.81) & (2.11) & (2.95) & (1.85) & (0.06) & (0.01) & (0.01)\\
\cline{2-9}
 & \multirow{2}{*}{Quartile} & 258.42 & 18.26 & 11.92 & 10.40 & 0.36 & 0.96 & 0.90\\

 &  & (5.22) & (3.46) & (4.68) & (3.19) & (0.11) & (0.02) & (0.02)\\
\cline{2-9}

 & \multirow{2}{*}{HB} & 231.02 & 5.66 & 39.32 & 23.00 & 0.80 & 0.85 & 0.85\\

 &  & (8.01) & (2.05) & (7.71) & (2.59) & (0.07) & (0.03) & (0.02)\\
\cline{2-9}
 & \multirow{2}{*}{RF} & 258.76 & 18.24 & 11.58 & 10.42 & 0.36 & 0.96 & 0.90\\
 &  & (5.28) & (3.19) & (4.71) & (2.99) & (0.10) & (0.02) & (0.02)\\
\cline{2-9}
 & \multirow{2}{*}{Tukey} & 248.98 & 10.88 & 21.36 & 17.78 & 0.62 & 0.92 & 0.89\\

 &  & (6.68) & (3.07) & (5.92) & (3.44) & (0.11) & (0.02) & (0.02)\\
\cline{2-9}
 & \multirow{2}{*}{Oracle} & 269.62 & 16.30 & 0.72 & 12.36 & 0.43 & 1.00 & 0.94\\

 &  & (1.85) & (2.37) & (0.70) & (2.23) & (0.07) & (0.00) & (0.01)\\
\hline
\multirow{14}{*}{Case (c)} & \multirow{2}{*}{Var} & 267.48 & 24.66 & 3.24 & 3.62 & 0.13 & 0.99 & 0.91\\

 &  & (2.70) & (2.48) & (2.40) & (1.94) & (0.07) & (0.01) & (0.01)\\
\cline{2-9}
 & \multirow{2}{*}{Const} & 267.00 & 26.12 & 3.72 & 2.16 & 0.08 & 0.99 & 0.90\\

 &  & (2.43) & (2.24) & (1.84) & (1.60) & (0.06) & (0.01) & (0.01)\\
\cline{2-9}
 & \multirow{2}{*}{Quartile} & 263.54 & 24.20 & 7.18 & 4.08 & 0.14 & 0.97 & 0.90\\

 &  & (3.47) & (2.83) & (2.90) & (2.27) & (0.08) & (0.01) & (0.01)\\
\cline{2-9}

 & \multirow{2}{*}{HB} & 225.94 & 12.32 & 44.78 & 15.96 & 0.56 & 0.83 & 0.81\\

 &  & (8.12) & (2.62) & (8.04) & (2.78) & (0.09) & (0.03) & (0.03)\\
\cline{2-9}
 & \multirow{2}{*}{RF} & 263.74 & 24.30 & 6.98 & 3.98 & 0.14 & 0.97 & 0.90\\

 &  & (3.33) & (2.78) & (2.90) & (2.23) & (0.08) & (0.01) & (0.01)\\
\cline{2-9}
 & \multirow{2}{*}{Tukey} & 250.84 & 18.68 & 19.88 & 9.60 & 0.34 & 0.93 & 0.87\\

 &  & (5.94) & (2.39) & (5.64) & (2.22) & (0.08) & (0.02) & (0.02)\\
\cline{2-9}
& \multirow{2}{*}{Oracle} & 269.78 & 23.52 & 0.94 & 4.76 & 0.17 & 1.00 & 0.92\\

 &  & (2.10) & (2.26) & (1.11) & (1.94) & (0.07) & (0.00) & (0.01)\\
\hline
\end{tabular}
\end{center}
\end{footnotesize} 
\caption{The performance measures of the five methods for the scenario with 5\% abnormal changes from 50 datasets. Numbers in parentheses are standard deviations. For each case, the variance function is as follows: 
(a) $\sigma^2(v_1, v_2) = 1$, (b) $\sigma^2(v_1, v_2) = (1/46) v_1^2$, and (c) $\sigma^2(v_1,v_2) = ({1}/{92}) (v_1+v_2)^2$.}
\label{tab:J5.revised}
\end{table}

\begin{table} [htb!]
\begin{footnotesize} 
\begin{center} 
\begin{tabular}{|c|c|c|c|c|c|c|c|c|}
\hline
Variance & Method & TN & FN & FP & TP & SEN & SPE & ACC\\
\hline
\multirow{14}{*}{Case (a)} & \multirow{2}{*}{Var} & 243.88 & 45.60 & 0.96 & 8.56 & 0.16 & 1.00 & 0.84\\
 &  & (3.43) & (3.77) & (1.23) & (3.23) & (0.06) & (0.01) & (0.01)\\
\cline{2-9}
 & \multirow{2}{*}{Const} & 244.22 & 45.84 & 0.62 & 8.32 & 0.15 & 1.00 & 0.84\\
 &  & (3.24) & (3.96) & (0.92) & (3.51) & (0.06) & (0.00) & (0.01)\\
\cline{2-9}
 & \multirow{2}{*}{Quartile} & 243.80 & 45.50 & 1.04 & 8.66 & 0.16 & 1.00 & 0.84\\
 &  & (3.56) & (4.02) & (1.56) & (3.64) & (0.06) & (0.01) & (0.01)\\
\cline{2-9}
 & \multirow{2}{*}{HB} & 220.86 & 19.26 & 23.98 & 34.90 & 0.64 & 0.90 & 0.86\\
 &  & (3.54) & (4.02) & (1.36) & (3.98) & (0.07) & (0.01) & (0.01)\\
\cline{2-9}
 & \multirow{2}{*}{RF} & 244.08 & 45.52 & 0.76 & 8.64 & 0.16 & 1.00 & 0.85\\
 &  & (3.09) & (3.15) & (0.00) & (0.71) & (0.01) & (0.00) & (0.01)\\
\cline{2-9}
 & \multirow{2}{*}{Tukey} & 233.90 & 26.98 & 10.94 & 27.18 & 0.50 & 0.96 & 0.87\\
 &  & (4.28) & (4.26) & (3.75) & (4.59) & (0.08) & (0.02) & (0.02)\\
\cline{2-9}
 & \multirow{2}{*}{Oracle} & 244.28 & 45.84 & 0.56 & 8.32 & 0.15 & 1.00 & 0.84\\
 &  & (3.19) & (3.41) & (0.70) & (2.66) & (0.05) & (0.00) & (0.01)\\
\hline
\multirow{14}{*}{Case (b)} & \multirow{2}{*}{Var} & 242.66 & 32.38 & 2.56 & 21.40 & 0.40 & 0.99 & 0.88\\
 &  & (4.26) & (5.17) & (2.70) & (5.06) & (0.09) & (0.01) & (0.02)\\
\cline{2-9}
 & \multirow{2}{*}{Const} & 241.28 & 46.40 & 3.94 & 7.38 & 0.14 & 0.98 & 0.83\\
 &  & (3.78) & (3.57) & (1.92) & (2.96) & (0.05) & (0.01) & (0.01)\\
\cline{2-9}
 & \multirow{2}{*}{Quartile} & 235.28 & 33.32 & 9.94 & 20.46 & 0.38 & 0.96 & 0.86\\
 &  & (5.14) & (5.18) & (3.78) & (5.03) & (0.09) & (0.02) & (0.02)\\
\cline{2-9}
& \multirow{2}{*}{HB} & 211.32 & 10.76 & 33.90 & 43.02 & 0.80 & 0.86 & 0.85\\
 &  & (7.69) & (2.82) & (7.12) & (3.77) & (0.05) & (0.03) & (0.02)\\
\cline{2-9}
 & \multirow{2}{*}{RF}  & 235.16 & 33.34 & 10.06 & 20.44 & 0.38 & 0.96 & 0.85\\
 &  & (5.37) & (5.15) & (4.05) & (5.10) & (0.09) & (0.02) & (0.02)\\
\cline{2-9}
 & \multirow{2}{*}{Tukey} & 225.36 & 18.92 & 19.86 & 34.86 & 0.65 & 0.92 & 0.87\\
 &  & (6.38) & (3.62) & (5.17) & (3.92) & (0.06) & (0.02) & (0.02)\\
\cline{2-9}
 & \multirow{2}{*}{Oracle} & 244.52 & 30.70 & 0.70 & 23.08 & 0.43 & 1.00 & 0.89\\
 &  & (3.36) & (4.13) & (0.79) & (4.05) & (0.07) & (0.00) & (0.01)\\
\hline
\multirow{14}{*}{Case (c)} & \multirow{2}{*}{Var} & 241.58 & 48.36 & 2.38 & 6.68 & 0.12 & 0.99 & 0.83\\
 &  & (3.49) & (3.79) & (2.78) & (3.57) & (0.06) & (0.01) & (0.01)\\
\cline{2-9}
 & \multirow{2}{*}{Const} & 240.84 & 51.28 & 3.12 & 3.76 & 0.07 & 0.99 & 0.82\\
 &  & (3.20) & (2.86) & (2.15) & (2.11) & (0.04) & (0.01) & (0.01)\\
\cline{2-9}
 & \multirow{2}{*}{Quartile} & 238.08 & 47.64 & 5.88 & 7.40 & 0.13 & 0.98 & 0.82\\
 &  & (3.92) & (3.85) & (3.36) & (3.42) & (0.06) & (0.01) & (0.01)\\
\cline{2-9}
 & \multirow{2}{*}{HB} & 206.10 & 22.90 & 37.86 & 32.14 & 0.58 & 0.84 & 0.80\\
 &  & (6.71) & (4.04) & (6.68) & (4.03) & (0.07) & (0.03) & (0.02)\\
\cline{2-9}
 & \multirow{2}{*}{RF} & 238.46 & 47.84 & 5.50 & 7.20 & 0.13 & 0.98 & 0.82\\
 &  & (3.84) & (3.82) & (3.35) & (3.36) & (0.06) & (0.01) & (0.01)\\
\cline{2-9}
 & \multirow{2}{*}{Tukey} & 227.20 & 36.38 & 16.76 & 18.66 & 0.34 & 0.93 & 0.82\\
 &  & (5.13) & (4.40) & (4.67) & (4.49) & (0.08) & (0.02) & (0.02)\\
\cline{2-9}
 & \multirow{2}{*}{Oracle} & 243.18 & 45.74 & 0.78 & 9.30 & 0.17 & 1.00 & 0.84\\
 &  & (2.93) & (3.00) & (0.89) & (2.97) & (0.05) & (0.00) & (0.01)\\
\hline
\end{tabular}
\end{center}
\end{footnotesize} 
\caption{The performance measures of the five methods for the scenario with 10\% abnormal changes from 50 datasets. Numbers in parentheses are standard deviations. For each case, the variance function is as follows: 
(a) $\sigma^2(v_1, v_2) = 1$, (b) $\sigma^2(v_1, v_2) = (1/46) v_1^2$, and (c) $\sigma^2(v_1,v_2) = ({1}/{92}) (v_1+v_2)^2$.}
\label{tab:J10.revised}
\end{table}

Tables 1 and 2 report the performance measures of the seven methods when the proportions of abnormal changes are $5\%$ and $10\%$, respectively. First, the tables show that our proposed method (Var) achieves a higher accuracy and smaller number of false positives than all other methods (Const, quartile, HB, RF, and Tukey) when the variance of the changes in the log(price) depends on the sales volume (cases (b) and (c)). Further, the performance measures of our method are close to those of Oracle, which uses the true variance function. Second, in case (a), the Const, quartile and RF methods and our method (Var) perform similarly better than the other two (Tukey and HB) methods in terms of accuracy. We remark that the Const, quartile, RF, and Var methods have the same interval under case (a). On the other hand, the HB method produces rather different results compared with the quartile method since the data generated by (\ref{eqn:simu-model}) are not close to one. Third, both the Tukey and HB methods have high sensitivity but low specificity in all three cases. This is because these methods tend to build narrower control limits than other methods, and the data points are more likely to be assigned as outliers.

\begin{figure} [h]
\centering
\begin{minipage}[b]{.3\textwidth}
\subfloat[]{\includegraphics[width=5.5cm,height=4.5cm]{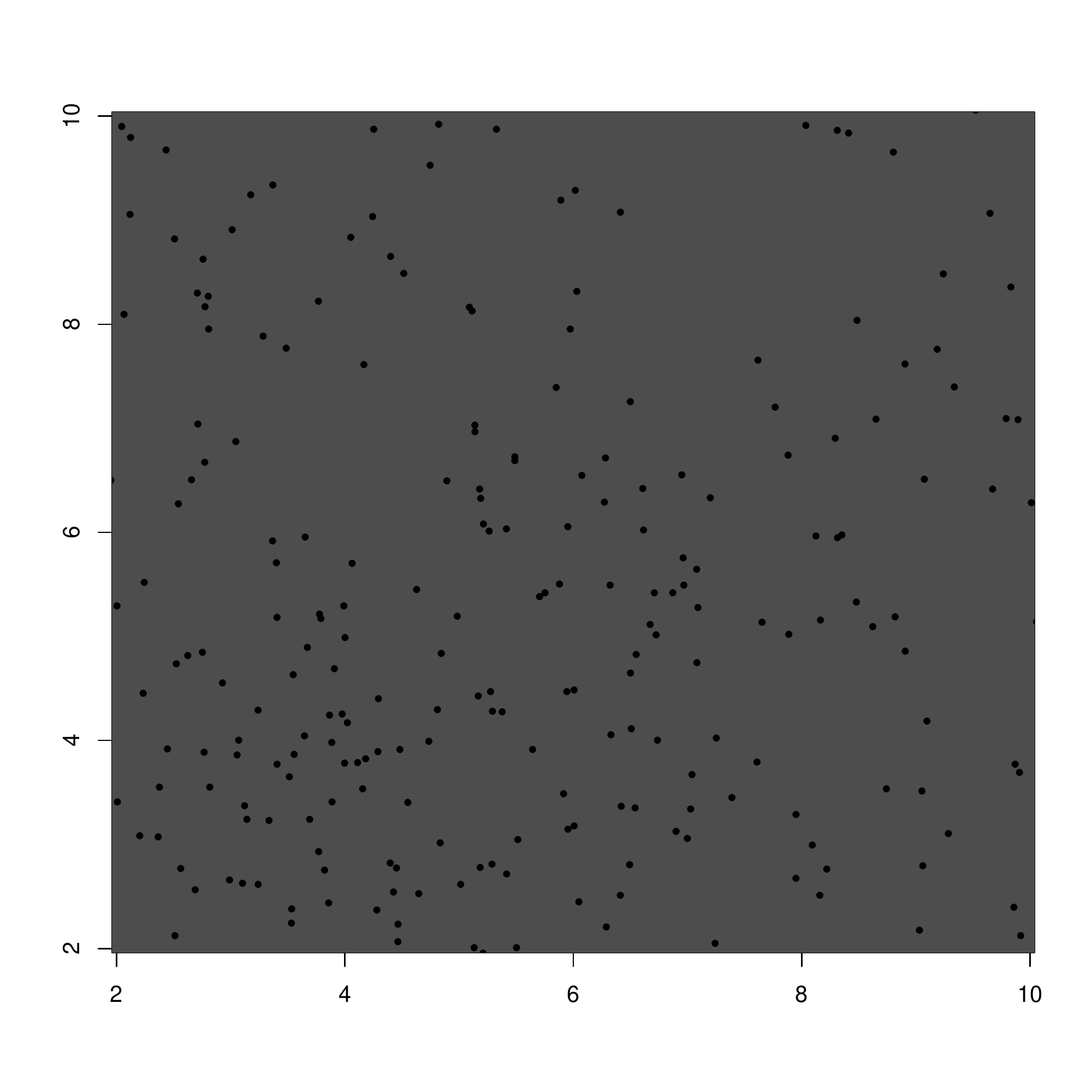}}
\end{minipage}\hfill
\begin{minipage}[b]{.3\textwidth}
\subfloat[]{\includegraphics[width=5.5cm,height=4.5cm]{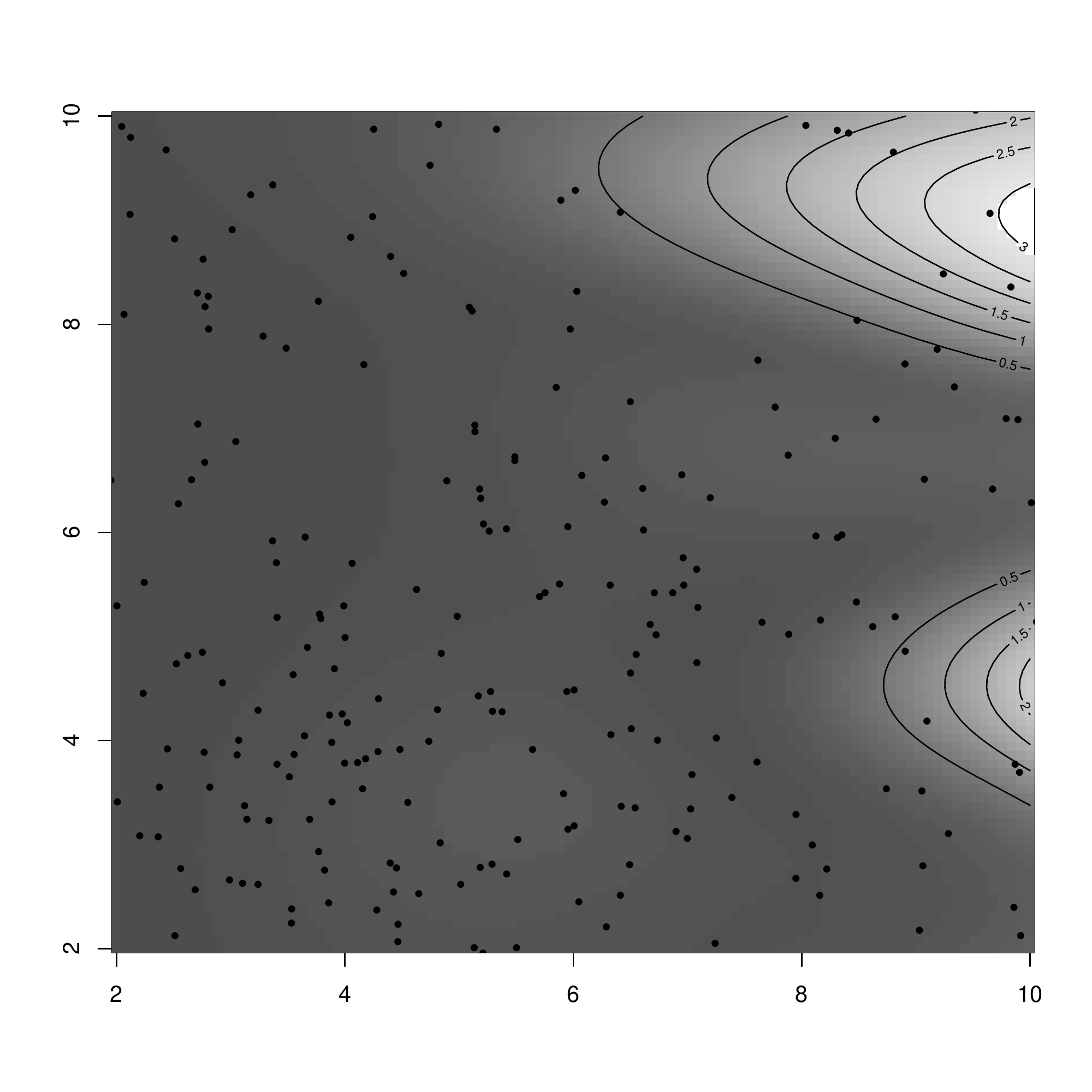}}
\end{minipage}\hfill
\begin{minipage}[b]{.3\textwidth}
\subfloat[]{\includegraphics[width=5.5cm,height=4.5cm]{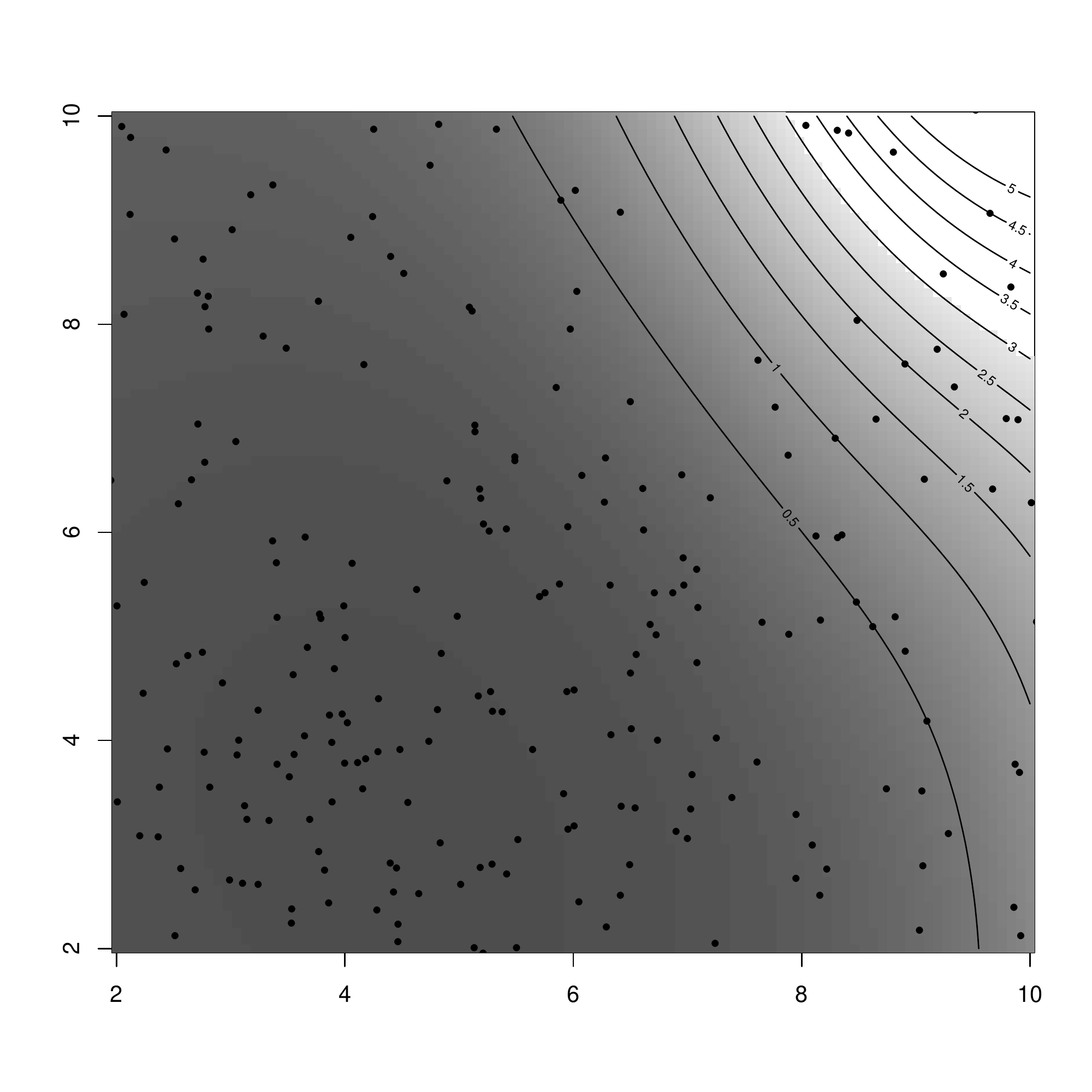}}
\end{minipage}
\caption{The mean square error of variance estimates with data points ($\cdot$). The left panel is for case (a), the center panel is for case (b), and the right panel is for case (c). Lighter gray represents larger variance estimates.}
\label{fig:msevar}
\end{figure}

We also investigate the accuracy of the variance estimation. Figure \ref{fig:msevar} displays the mean square error (MSE) of variance estimates for each case. In the figure, the black dots are the observed data points of $(V_{t-1},V _t)$. The figure shows that the MSE is relatively small for the area of $(v_1,v_2)$, where more data points are observed, whereas it is large on the upper-right side of $v_1$ and $v_2$, where few data points are observed.

\section{Data example}

In this section, we apply the proposed detection method to real scanner data obtained from the Korean Chamber of Commerce and Industry (KCCI). These scanner data were collected at weekly intervals between 2013 and 2014 from approximately 2,000 retail stores. We monitor price changes for one popular item, that is, A-brand cartons of milk for toddlers.
We set up a retail store and monitor potential anomalies in price changes for this item. We calculate the weekly average price, $P_t$, and the log of the price change $Y_t = \log (P_t / P_{t-1}) = \log (R_t )$, for $t=1, 2,  \cdots, T$. In this section, the Tukey samples, that is, those with all values of $R_t=1$ removed, are used for all methods to ensure a fair comparison. The portion of the samples with $R_t=1$ is more than 90\%, and the quartile, HB and RF methods perform extremely poorly when the original samples are used.

The data collected in 2013 in all of the stores are used as the training samples to compute control limits, and the data collected in 2014 for the specified store are used as the test samples for monitoring.

\begin{figure}[htp!]
\centering
\begin{minipage}[b]{.5\textwidth}
\subfloat[]{\includegraphics[width=8cm,height=8cm]{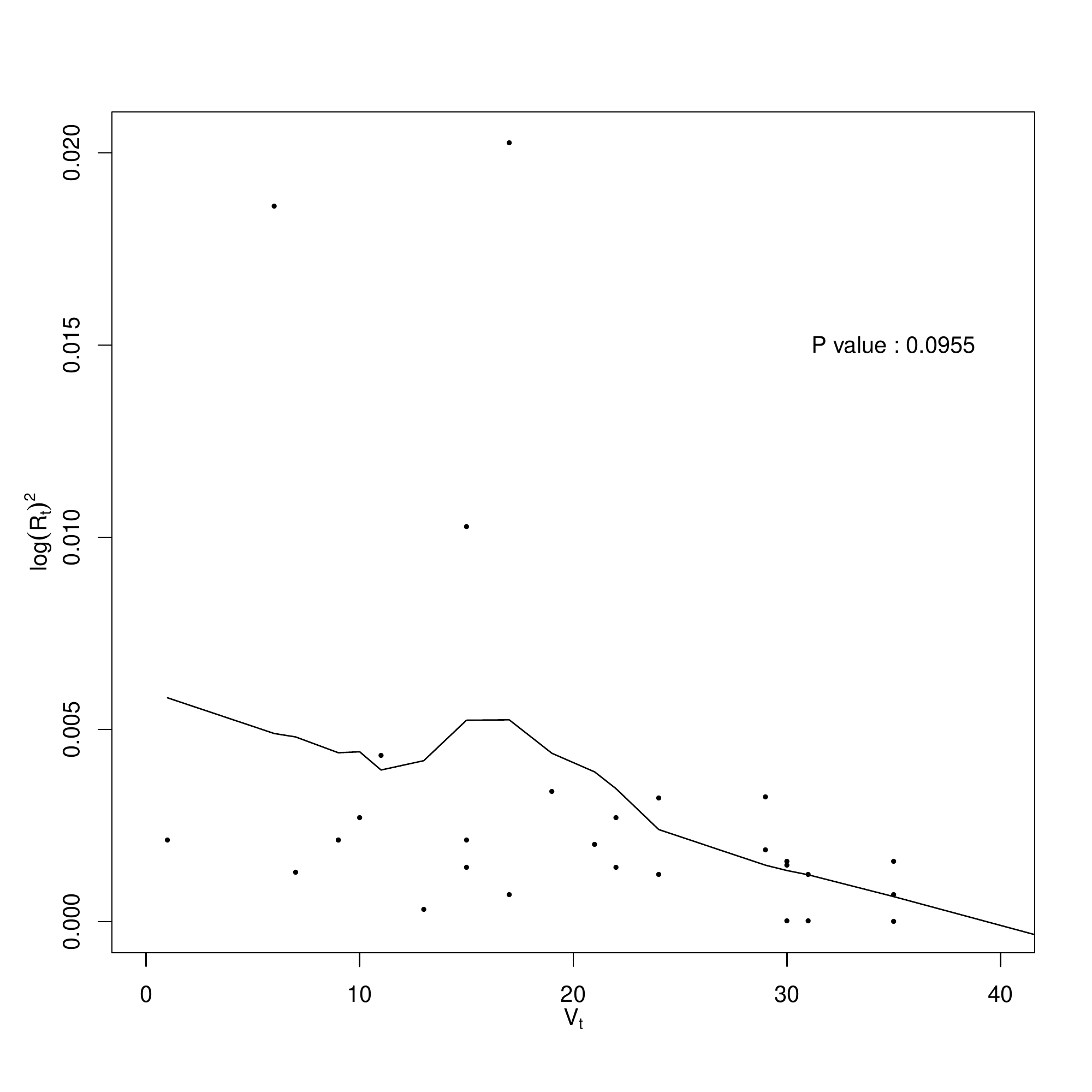}}
\end{minipage}\hfill
\begin{minipage}[b]{.5\textwidth}
\subfloat[]{\includegraphics[width=8cm,height=8cm]{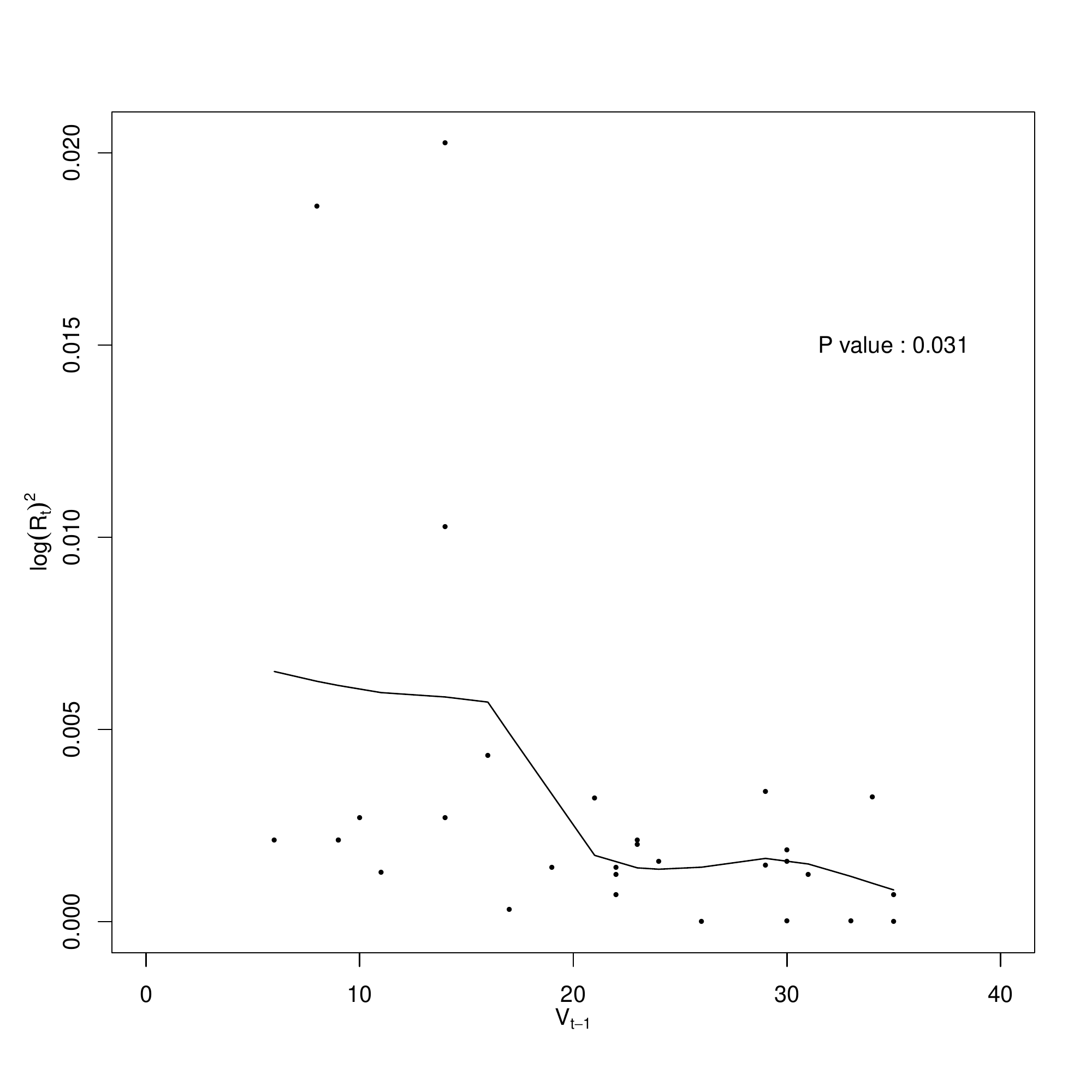}}
\end{minipage}\hfill
\caption{ Plots of (a) $y_t^2$ versus sales volume at time $t$, $v_t$, and (b) $y_t^2$ versus sales volume at time $t-1$, $v_{t-1}$ for the price of the `A-brand cartons of milk for toddlers' product from the KCCI. The $p-$values are obtained from the chi-square test for comparing the local constant model with the constant model. In the local constant model, the ratio of data used in the estimation is 50\%, which is a default in a gamma function.}
\label{fig:y2_vs_v}
\end{figure}

Before we apply the methods, to show whether the variance of the log of the price change is affected by sales volume in the training samples, we visualize the relationship of $Y^2_t$ with $V_t$ and $V_{t-1}$, respectively, applying the generalized additive model.
We plot the figures using the `{\it gam}' function of the `{\it gam}' package. Figure \ref{fig:y2_vs_v} shows these two plots ($Y^2_t$, $V_t$) and ($Y^2_t$, $V_{t-1}$), which reveal a decreasing pattern with increasing sales volume. This pattern indicates a problem with the specification of a constant variance for $Y_t$. The smaller p-values suggest that the specification contradicts the model with constant variance.

\begin{figure}[htb!]
\begin{center}
\includegraphics[width=15cm, height=9cm]{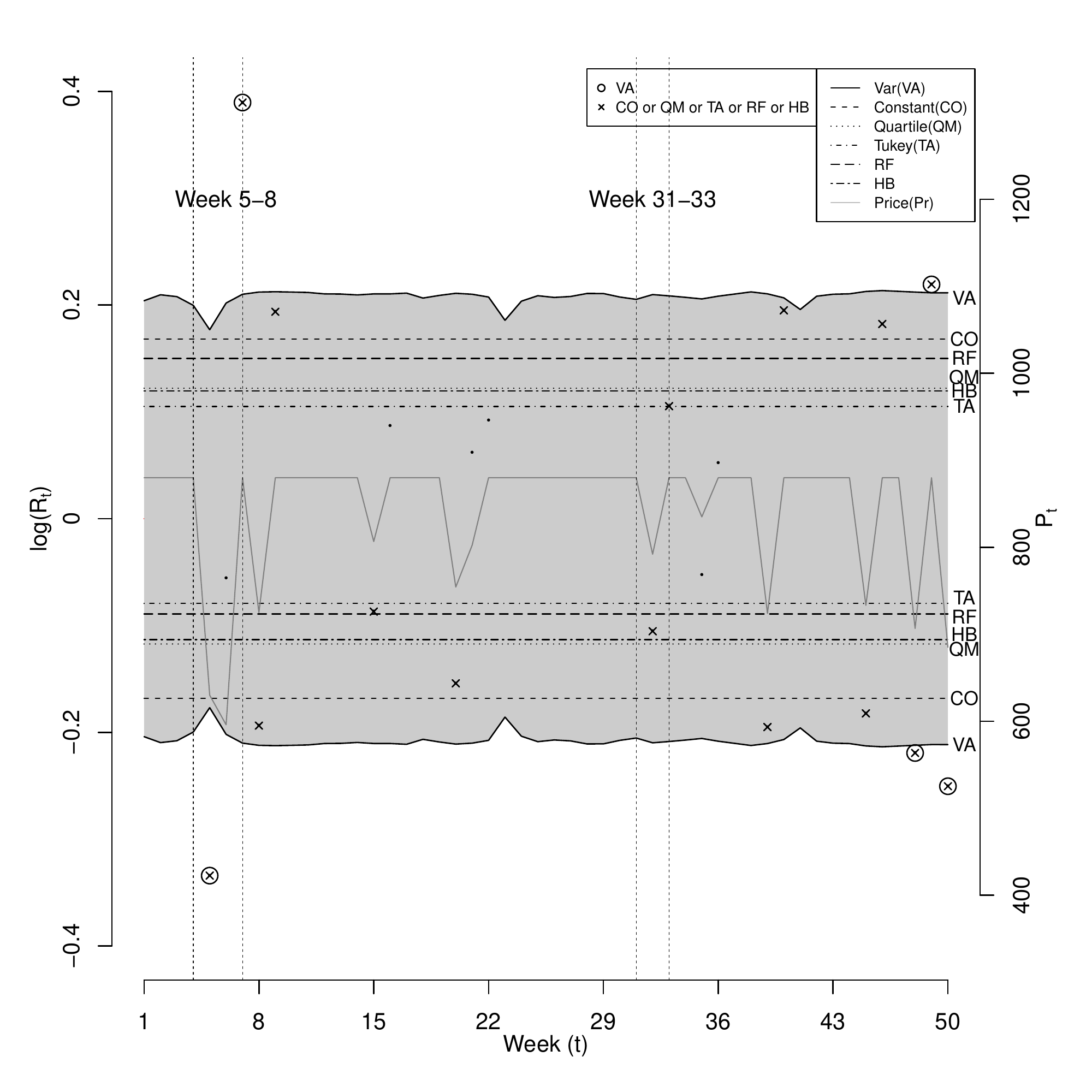}
\caption{Anomaly monitoring for the `A-brand cartons of milk for toddlers' product with various anomaly detection methods. The y-axes on the left and the right represent the log of the price change, $\log (R_t)$, and the weekly average price, $P_t$ (unit: KRW), respectively.}
\label{fig:chart}
\end{center}
\end{figure}

We aim to compare six different methods (Var, Const, quartile, HB, RF, Tukey) to establish control limits for monitoring price changes as in the simulation study. Figure \ref{fig:chart} shows the monitored results for the log of the price change in the test sample, with six control limits from six methods computed by the training samples. The x-axis represents the weeks that are observed, and the observations are made weekly. The y-axis on the left represents the log of the price change, $\log (R_t)$. To facilitate the understanding of the original price, the y-axis on the right represents the weekly average price, $P_t$ (unit: KRW). The solid lines show the control limit of the proposed method (Var), and dashed, dotted, dot-dashed, long-dashed, and two-dashed lines correspond to the Const, quartile, Tukey, RF and HB methods, respectively. The gray solid path in the middle represents the weekly average price. In addition, `$\cdot$' represents the log of the price change, `$\circ$' represents the outlier detected by the variable method, and `$\times$' represents the outlier detected by the existing methods (Const, quartile, HB, RF and Tukey). We recognize that our new method identifies fewer rare events than all of the other methods, which is consistent with the simulation study. We also find that the interval of Const is wider than those of other methods (quartile, HB, and RF).

In Figure \ref{fig:chart}, $15$ outliers are found by all of the existing methods, but only $5$ outliers are found by the variable method, implying that the latter method may tend to show high specificity. From the observations between weeks $5$ and $8$, two points are judged as outliers in all methods, and we can see that these are cases when the price dropped from KRW $880$ to KRW $630$ and again increased to KRW $880$. It is not surprising that these price changes are detected since the prices change by approximately $28.4\%$.

On the other hand, from the observations between weeks $31$ and $33$, the Tukey method detects the $32$nd and $33$rd points as outliers while the Var method does not. The price changes around week $32$ are relatively small, approximately KRW $100$ in width. This means that when the price change is regarded as sufficiently acceptable, the new method in this study does not judge the observations as obvious outliers or unusual observations. This is a more reasonable judgment for predicting outliers.

\begin{figure}[htb!]
\begin{center}
\includegraphics[width=12cm, height=7.5cm]{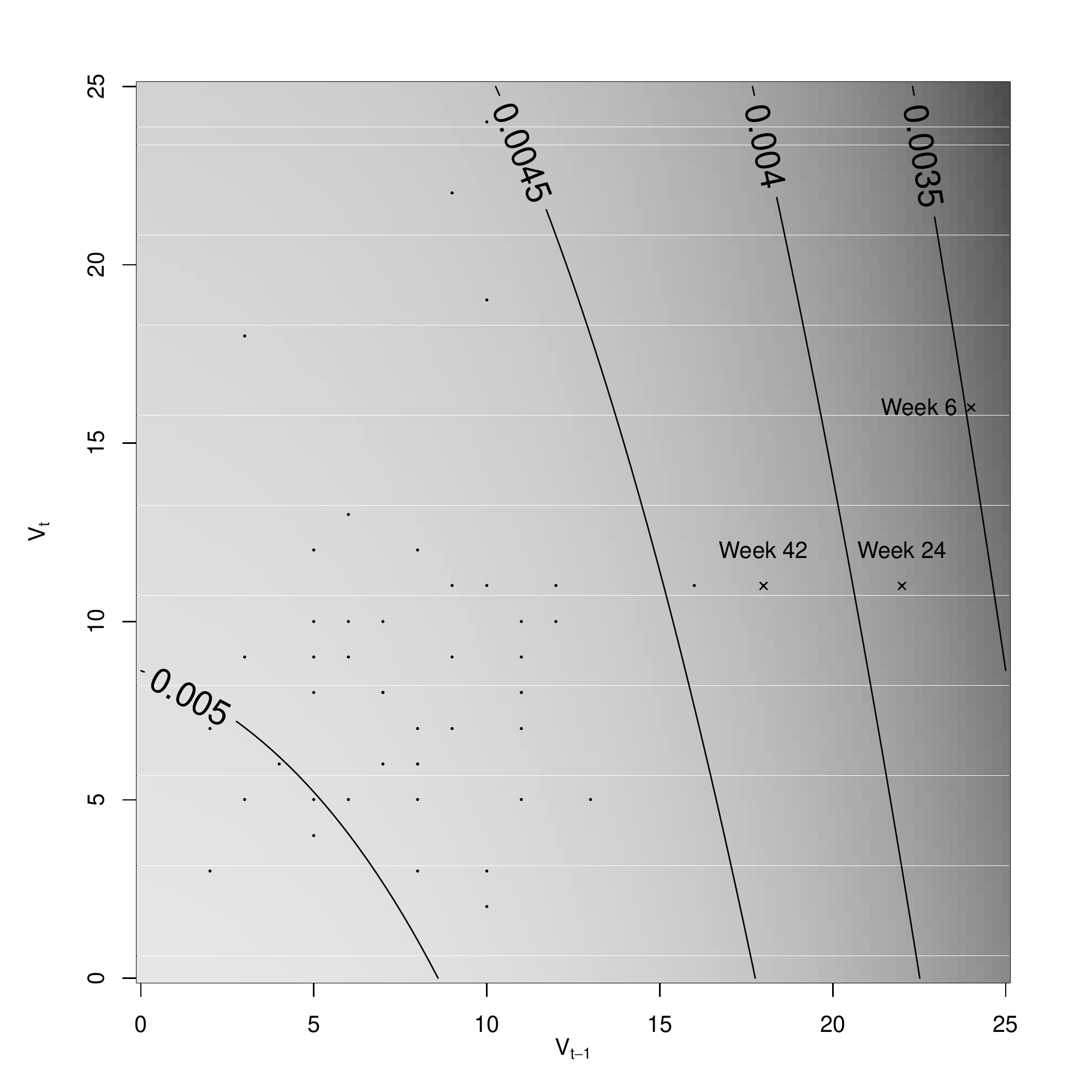}
\caption{Variance estimates of the log of the price change rate of the `A-brand cartons of milk for toddlers' product. Lighter gray represents larger variance estimates. `$\cdot$' and `$\times$' mark observed $(V_{t-1}, V_t)$ in the test samples.}
 \label{fig:hist}
\end{center}
\end{figure}

Figure \ref{fig:hist} shows the contour plot of the variance estimates for the log of price changes of this item as a function of sales volume at times $t-1$ and $t$. The x-axis and the y-axis indicate the sales volume of the test sample at times $t-1$ and $t$, respectively. Note that we choose one retail store as a test sample, where the weekly average of the sales volume lies between $0$ and $25$. In Figure \ref{fig:hist}, the variance estimates when the sales volume is relatively small (bottom left) appear larger than the estimates when the sales volume is relatively large (top right). In addition, particularly in weeks 6, 24, and 42, the control limits of the Var method appear narrower than those of the other methods, as shown in Figure \ref{fig:chart}. Consequently, the variance estimates are small.

These characteristics allow for conservative anomaly detection when the price change varies depending on the sales volume, i.e., when the variables of interest are affected by the information of other explanatory variables.

\section{Conclusion}

In this study, we propose an outlier detection method based on the fact that the variance of the price change depends on the sales volume. While existing methods judge whether price changes are detected irrespective of sales volume, the proposed method (Var) reflects the fact that the dispersion of price changes can be influenced by another observed covariate (sales volume). This study is motivated by the utility of employing scanner data collected by the KCCI; other applications of our method could easily be found in, for example, health-care monitoring or public-health surveillance \citep{Wodall:2006}. See, for example, \citet{Cover:2011}. To build the proper control limits, we model the variance of the log values of the price change that is under control as a smooth function of the sales volume, and we adopt the local polynomial regression to estimate the function. The simulation results and empirical analysis show the advantages of the new proposal over existing methods on various measures of performance, especially its high accuracy, reduced number of false positives and, hence, increased specificity. 

We now conclude the paper with a discussion on the covariate-dependent mean of the log of the price. If the mean of the log of the price that is under control also depends on the sales volume, the log of the price is modeled as follows:
\begin{equation}\label{eqn:model2}
Y_t =\mu(V_{t-1}, V_t) +\sigma(V_{t-1}, V_t) \epsilon_t,
\end{equation}
where $\mu(x_1, x_2)$ and $\sigma(x_1,x_2)$ are continuous functions in $x_1$ and $x_2$, and all other variables are defined the same as those in (\ref{eqn:model}). 
In the sequel, the control limit with type I error $\alpha$ at time $s$ becomes
\begin{equation} \label{eqn:oracle-limits}
\widehat{\mu}(V_{s-1},V_s\big) \pm 3 \widehat{\sigma}(V_{s-1}, V_{s}),
\end{equation}
where $V_{s}$ and $V_{s-1}$ are the sales volumes at times $s$ and $s-1$, respectively, and $\widehat{\mu}$ and $\widehat{\sigma}$
are the smooth mean and variance function, respectively. However, in our analysis, we consider the price under control as a constant that does not depend on the sales volume because, otherwise, it raises the fundamental question, ``What is the price of a product?".

\vskip1cm
\bibliographystyle{apalike}

\begin{thebibliography}{}


\bibitem[Bird et al., 2014]{Bird:2014} Bird, D., Breton, R., Payne, C., and Restieaux, A. (2014). Initial report on experiences with scanner data in ONS, Office for National Statistics, UK. \url{http://www.ons.gov.uk/ons/guide-method/usesr-guidance/prices/cpi-and-rpi/intial-report-on-experiences-with-scanner-data-in-ons.pdf} (accessed 17 Feburary, 2019).

\bibitem[Haan and van der Grient, 2011]{deHaan:2011} Haan, J.de. and van der Grient, H. (2011). Eliminating chain drift in price indexes based on scanner data, \emph{Journal of Econometrics}, {\bf 161}, 36-46.

\bibitem[Haan and Krsinich, 2014]{deHaan:2014} Haan, J.de. and Krsinich, F. (2014). Scanner data and the treatment of quality change in nonrevisable price indexes, \emph {Journal of Business \& Economic Statistics}, {\bf 32}, 341-358.


\bibitem[Saidi and Rubin-Bleuer, 2005]{Saidi:2005} Saidi, S. and Rubin-Bleuer, S. (2005). Detection of outliers in the canadian consumer price index, \emph{Business Survey Methods Division, Statistics Canada}, {\bf 5}, 16-18.

\bibitem[Rais, 2008]{Rais:2008} Rais, S. (2008). Outlier detection for the consumer price index, \emph{Proceeding of Statistical Society of Canada}. \url{http://ssc.ca/sites/default/files/survey/documents/SSC2008_S_Rais.pdf} (accessed 17 Feburary, 2019).

\bibitem[Tukey, 1977]{Tukey:1977} Tukey, J. W. (1977). \emph{Exploratory data analysis}, Addison-Wesley, Massachusetts.


\bibitem[ILO Consumer Price Index Manual, 2004]{ILO:2004} International Labor Organization, International Monetary Fund, Organization for Economic Co-operation and Development, United Nations Economic Commission for Europe, The World Bank (2004). \emph{Consumer Price Index Manual: Theory and Practice}. \url{https://www.ilo.org/wcmsp5/groups/public/---dgreports/---stat/documents/presentation/wcms_331153.pdf} (accessed 17 Feburary, 2019).

\bibitem[Hall and Carroll(1989)]{HC:89} Hall, P. and Carroll, R. J. (1989). Variance function estimation in regression: the effect of estimating the mean. \emph{Journal of the Royal Statistical Society - Series B}, {\bf 1}, 3-14.


\bibitem[Fan and Yao, 1998]{FY:98} Fan, J. and Yao, Q. (1998). Efficient estimation of conditional variance functions in stochastic interest rates, \emph{Biometrika}, {\bf 85}, 645-660.


\bibitem[Brown and Levine, 2007]{brown2007variance} Brown, L. D. and Levine, M. (2007). Variance estimation in nonparametric regression via the difference sequence method. \emph{The Annals of Statistics}, {\bf 35(5)}, 2219-2232.


\bibitem[Thompson and Sigman, 1999]{Thompson:1999} Thompson, K. and Sigman, S. (1999). Statistical methods for developing ratio edit tolerances for economic data, \emph{Journal of Official Statistics}, {\bf 15(4)}, 517-535.

\bibitem[Hidiroglou and Berthelot, 1986]{Hidi:1986} Hidiroglou, M. A. and Berthelot, J. M.  Statistical editing and imputation for periodic business surveys, \emph{Survey Methodology}, {\bf 12}, 73-84.

\bibitem[Office for National Statistics, 2014]{ONS:2014} Office for National Statistics (2014). Consumer price indices technical manual, UK. \url{http://doc.ukdataservice.ac.uk/doc/7022/mrdoc/pdf/7222technical_manual_2014.pdf} (accessed 17 Feburary, 2019).

\bibitem[Wodall, 2006]{Wodall:2006} Woodall, W.H. (2006). The use of control charts in health-care monitoring and public-health surveillance, \emph{Journal of quality technology}, {\bf 38}(2), 89-104.

\bibitem[Cover and Schopflocher, 2011]{Cover:2011} Cover, D.C. and Schopflocher, D.P. (2011). Using funnel plots in public health surveillance, \emph{Population health metrics}, {\bf 9}:58,  doi:10.1186/1478-7954-9-58.


\end{thebibliography}

\end{document}